\documentclass[a4paper]{aa}
\usepackage[T1]{fontenc}
\usepackage[utf8]{inputenc}
\usepackage{graphicx}
\usepackage{color}
\usepackage{txfonts}
\usepackage{microtype}
\usepackage{ellipsis}
\usepackage{natbib}

\bibpunct{(}{)}{;}{a}{}{,}
\usepackage[pdftex,colorlinks=true,urlcolor=blue]{hyperref}
\usepackage[xindy,nonumberlist,nopostdot,nogroupskip]{glossaries}

\makeglossaries
\usepackage[xindy]{imakeidx}
\makeindex

\newcommand{\capurl}[1]{{\fontsize{8}{10}\selectfont\url{#1}}}
\newcommand{\action}[1]{
{\medskip\noindent
\bfseries
\begin{tabular}{p{0.9cm}p{7.3cm}}
Action: & #1\\
\end{tabular}

\medskip
}
}

\begin{document}
\title{A View From Above}
\author{M. Nielbock}
\institute{Haus der Astronomie, Campus MPIA, Königstuhl 17, D-69117 
Heidelberg, Germany\\
\email{nielbock@hda-hd.de}}

\date{Received February 18, 2016; accepted }

\abstract{This activity has been developed as a resource for the ``EU Space Awareness'' educational programme. As part of the suite ``Our Fragile Planet'' together with the ``Climate Box'' it addresses aspects of weather phenomena, the Earth's climate and climate change as well as Earth observation efforts like in the European ``Copernicus'' programme. In this activity, students investigate how satellite images obtained at different wavelengths help to identify Earth surface features like vegetation and open water areas by using a specially designed software package, LEO Works.  Students inspect and analyse real satellite data to produce colour images and maps of spectral indices and learn how to interpret them and their uses.}


\maketitle
%
%________________________________________________________________

\section{Background information}
\subsection{Remote sensing}
The term
\newglossaryentry{remote}{name={Remote sensing},description={A measurement technique that probes and analyses the Earth from outer space, via satellites.}}
{\em remote sensing} indicates a measurement technique that probes and analyses the Earth from outer space. Alongside classical
\newglossaryentry{insitu}{name=in-situ,description={This is an expression derived from Latin which means measuring a phenomenon directly where it appears in Nature. It can be regarded as the complement of remote sensing.}}
in-situ methods like weather stations, field surveying or taking samples, satellite based measurements are becoming an increasingly important source of data. The advantage is the fast and complete coverage of large areas. However, satellite data are not always easy to interpret and need substantial treatment.

The most abundant remote sensing devices are weather satellites. By employing suitable sensors, they provide information about cloud coverage, temperature distributions, wind speed and directions, water levels and snow thickness. Keeping the evolving climate change in mind, those data play an increasing important role in disaster management during draughts and floods, climate simulations, atmospheric gas content and vegetation monitoring. In addition, urban and landscape management benefit from satellite data.

\begin{figure}
\centering
\resizebox{0.6\hsize}{!}{\includegraphics{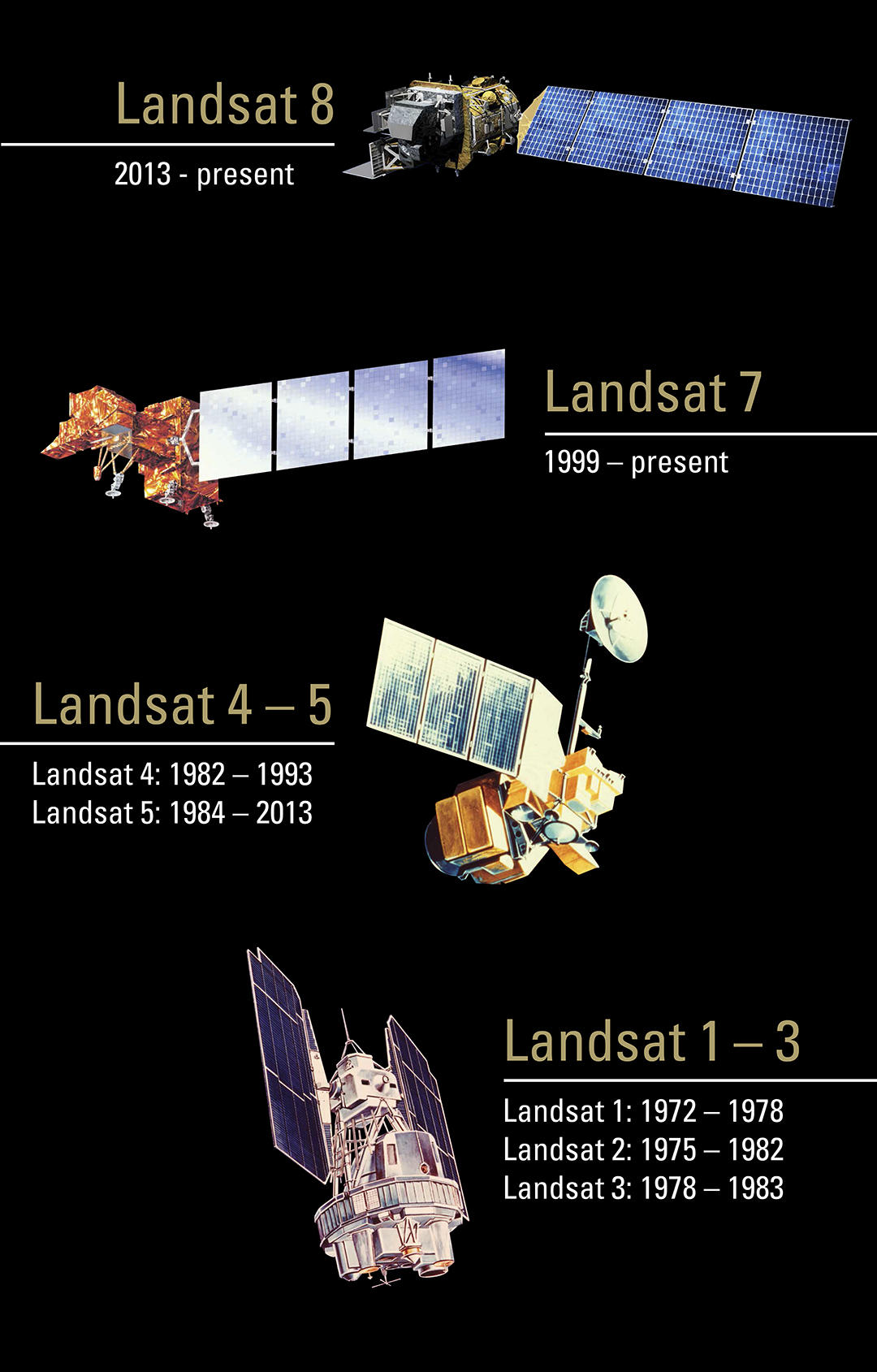}}
\caption{Overview of Landsat remote sensing satellites of NASA (NASA, \capurl{https://www.usgs.gov/media/images/landsat-program}).}
\label{f:landsat}
\end{figure}

The first weather satellites were launched by NASA as early as 1960. In the beginning of the 1970s, NASA started their Earth observation programme using Landsat satellites (Fig.~\ref{f:landsat}). In Europe, France was first using their SPOT satellite fleet. They were followed by the remaining European countries in the 1990s after the foundation of ESA, the European Space Agency.

\subsection{The Copernicus Programme}
\begin{figure}
\centering
\resizebox{\hsize}{!}{\includegraphics{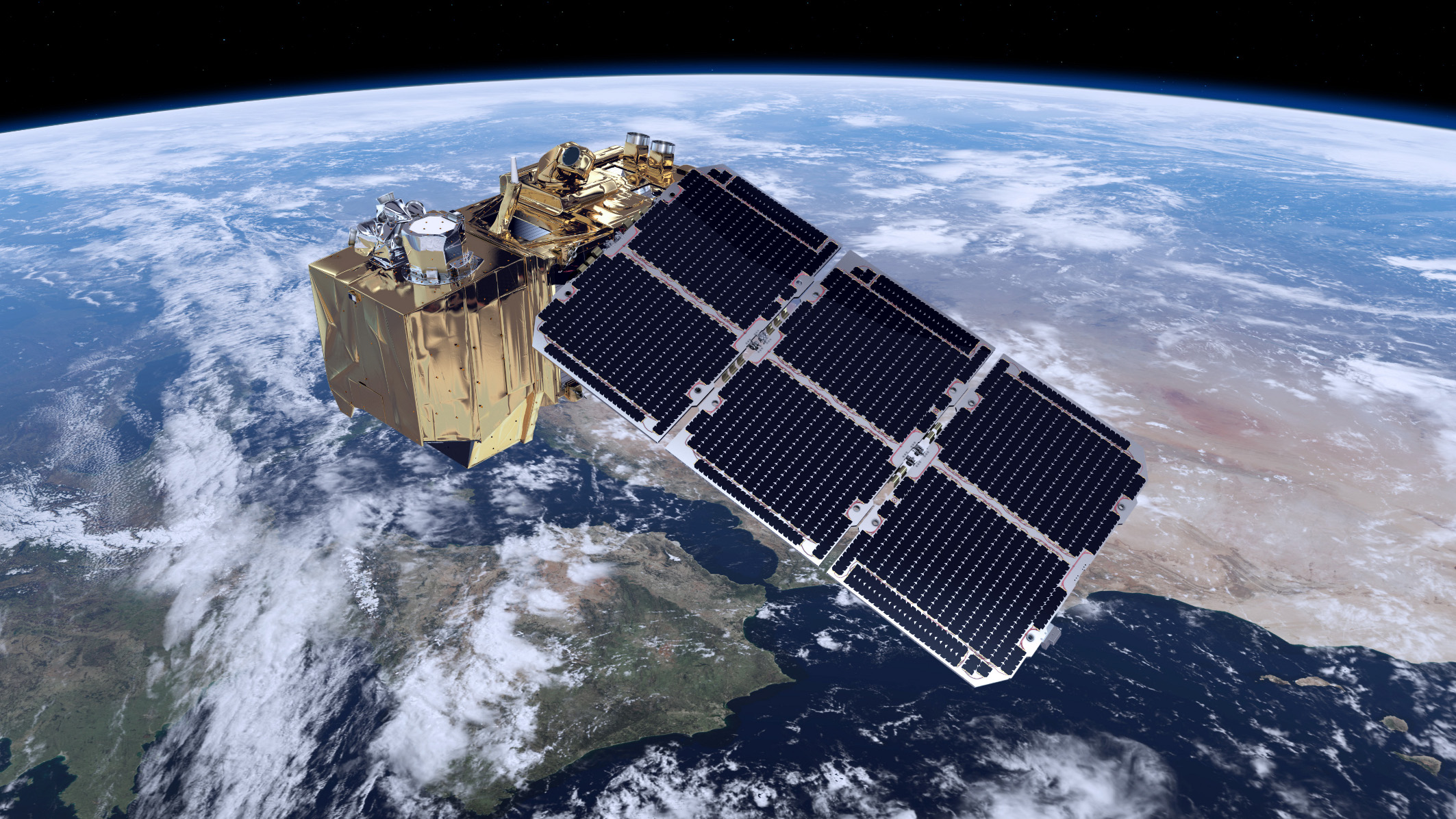}}
\caption{Computer model of the Sentinel-2A satellite launched on 23 June 2015 (ESA/ATG medialab, \capurl{http://www.esa.int/spaceinimages/Images/2014/07/Sentinel-2_brings_land_into_focus}).}
\label{f:sentinel2a}
\end{figure}

Already since 1997 the USA and NASA have been building a large programme for exploring the Earth, labelled the Earth Observation System, which consist of a large number of different satellites. Starting in 1998, the European equivalent, the Global Monitoring for Environment and Security (GMES) is being developed. In 2012, the programme was renamed to Copernicus\footnote{\url{http://www.copernicus.eu}}. Information products for six applications are being derived from the satellite data: ocean, land and atmosphere monitoring, emergency response, security and climate change. The data products are offered to everyone free of charge. They are supplied via two branches: space based remote sensing devices (satellite component) as well as airborne, ground and marine probing (in-situ component). The core of the satellite component is the fleet of Sentinel satellites that have been and are being built exclusively for the Copernicus projects. They are supplemented by other domestic and commercial partner missions. The first Sentinel satellite (Sentinel 1-A) was launched in 2014. Sentinel-2A (Fig.~\ref{f:sentinel2a}) and 3-A followed 2015 and 2016, respectively.

\subsection{Electromagnetic spectrum}
\begin{figure}[!ht]
\centering
\resizebox{\hsize}{!}{\includegraphics{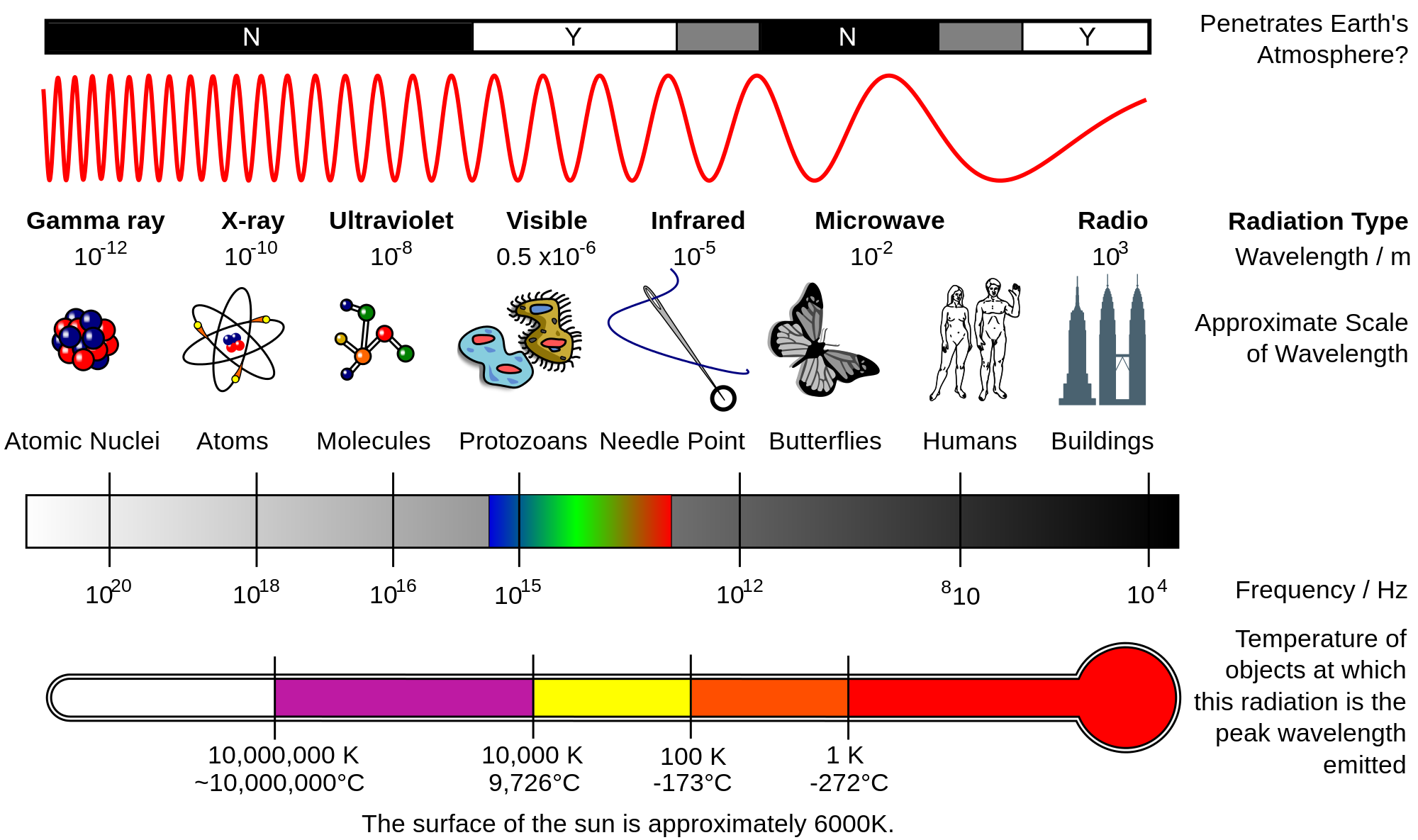}}
\caption{The spectrum of electromagnetic radiation. The visible light is only a very small part inside the full range (Inductiveload, \capurl{https://commons.wikimedia.org/wiki/File:EM_Spectrum_Properties_reflected.svg}, ``EM Spectrum Properties reflected'', cropped by Markus Nielbock, \capurl{https://creativecommons.org/licenses/by-sa/3.0/legalcode}).}
\label{f:emspectrum}
\end{figure}

The kind of radiation that the human eye can see and interpret is called light. However, the full range of electromagnetic radiation (the spectrum) is much bigger. The part that is invisible to us can be detected by special cameras, such as the ones put on astronomical telescopes and satellites. A good overview on the different kinds of radiation is provided in Fig.~\ref{f:emspectrum}.

\subsection{Multispectral imaging}
One of the core purposes of Earth observation and remote sensing is taking and analysing pictures. Similar to modern astronomy, taking images with different spectral filters is very diagnostic when identifying and analysing terrestrial surface features. For this kind of data acquisition, the cameras rely on the sunlight that illuminates the Earth's surface. Hence, they receive the portion of the sunlight that is reflected by the various surface features. Compared to the incident sunlight, the reflected light is modified by brightness and spectral composition.

\begin{table}
\caption{Spectral bands of the MSI camera of the Sentinel-2A satellite (Sentinel Online, \capurl{https://sentinels.copernicus.eu/
web/sentinel/technical-guides/sentinel-2-msi/msi-instrument}).}
\label{t:s2specbands}
\begin{tabular}{cccc}
\hline\hline
Band & Central & Bandwidth & Spatial \\
     & wavelength ($\mu$m) & ($\mu$m) & resolution (m) \\
\hline
1	& 0.443	& 0.020	& 60\\
2	& 0.490	& 0.065	& 10\\
3	& 0.560	& 0.035	& 10\\
4	& 0.665	& 0.030	& 10\\
5	& 0.705	& 0.015	& 20\\
6	& 0.740	& 0.015	& 20\\
7	& 0.783	& 0.020	& 20\\
8	& 0.842	& 0.115	& 10\\
8a	& 0.865	& 0.020	& 20\\
9	& 0.945	& 0.020	& 60\\
10	& 1.380	& 0.030	& 60\\
11	& 1.610	& 0.090	& 20\\
12	& 2.190	& 0.180	& 20\\
\hline
\end{tabular}
\end{table}

\begin{figure}[!ht]
\centering
\resizebox{\hsize}{!}{\includegraphics{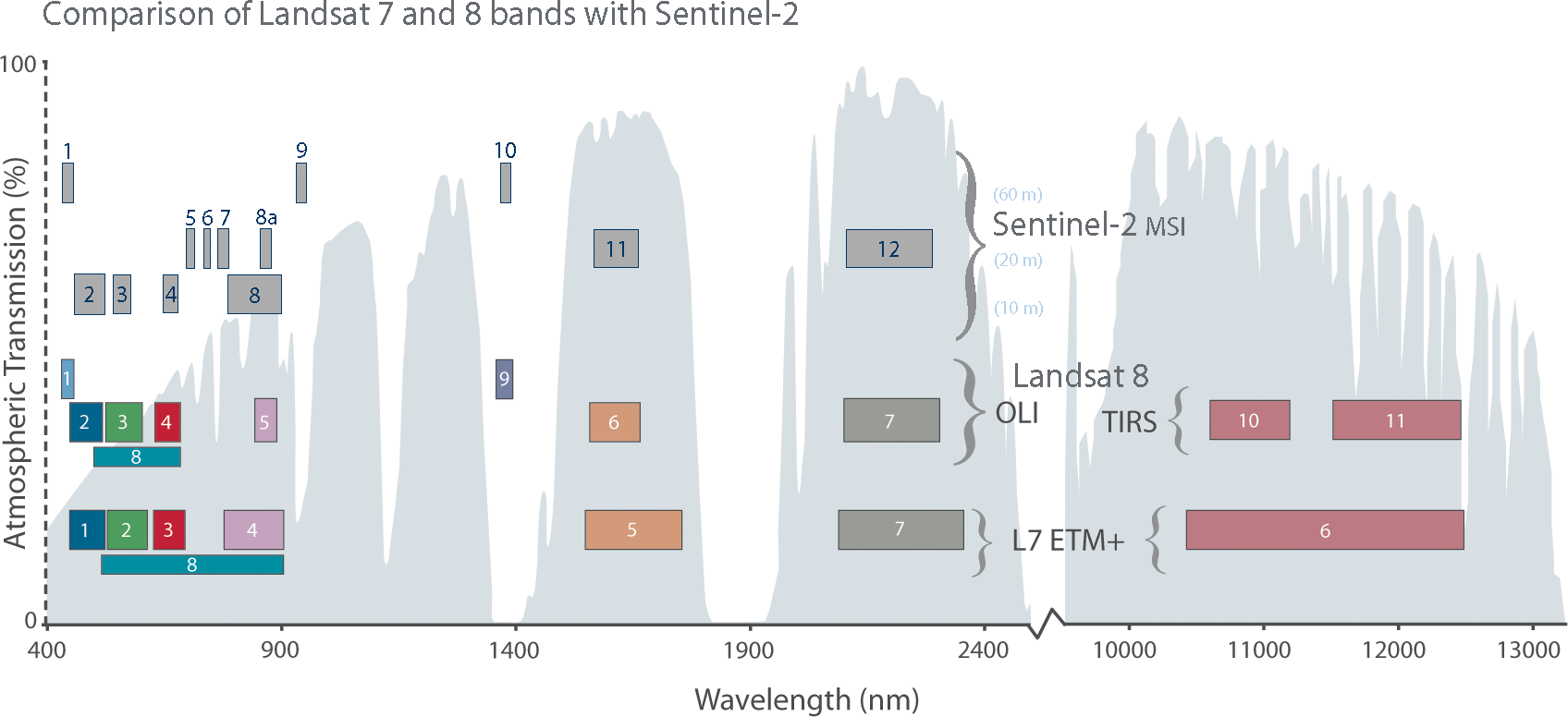}}
\caption{Graphical representation of the spectral bands of MSI/Sentinel-2A compared to the cameras of the Landsat 7 and 8 satellites. The axes depict the wavelength in nanometres ($1\,{\rm nm} = 10^{-3}\,\mu{\rm m} = 10^{-9}\,{\rm}$) and the terrestrial atmospheric transmission (grey) in percent (NASA, \capurl{https://landsat.gsfc.nasa.gov/sentinel-2a-launches-our-compliments-our-complements/}).}
\label{f:transm}
\end{figure}

The spectral bands of the camera ``Multi-Spectral Instrument (MSI)'' of the Sentinel-2A satellite is given as an example in Tab.~\ref{t:s2specbands}. For example, band 2 covers a wavelength range of $0.065\,\mu$m centred on a wavelength of $0.490\,\mu$m. The smallest feature that could be seen in this band would be 10~m across. Those bands cannot be chosen arbitrarily because of the wavelength dependent transparency of the Earth's atmosphere (grey area in Fig.~\ref{f:transm}). They are referred to as spectral windows. The main culprit for the wavelength ranges, where the atmosphere blocks external radiation, is water vapour. Therefore, observations with cameras have to be designed in a way that only those wave bands are used, where the radiation is transmitted well enough to receive a good signal. Thus, these ranges are the ones the optical filters of the cameras are designed for.

\begin{figure*}
\centering
\resizebox{\hsize}{!}{\includegraphics{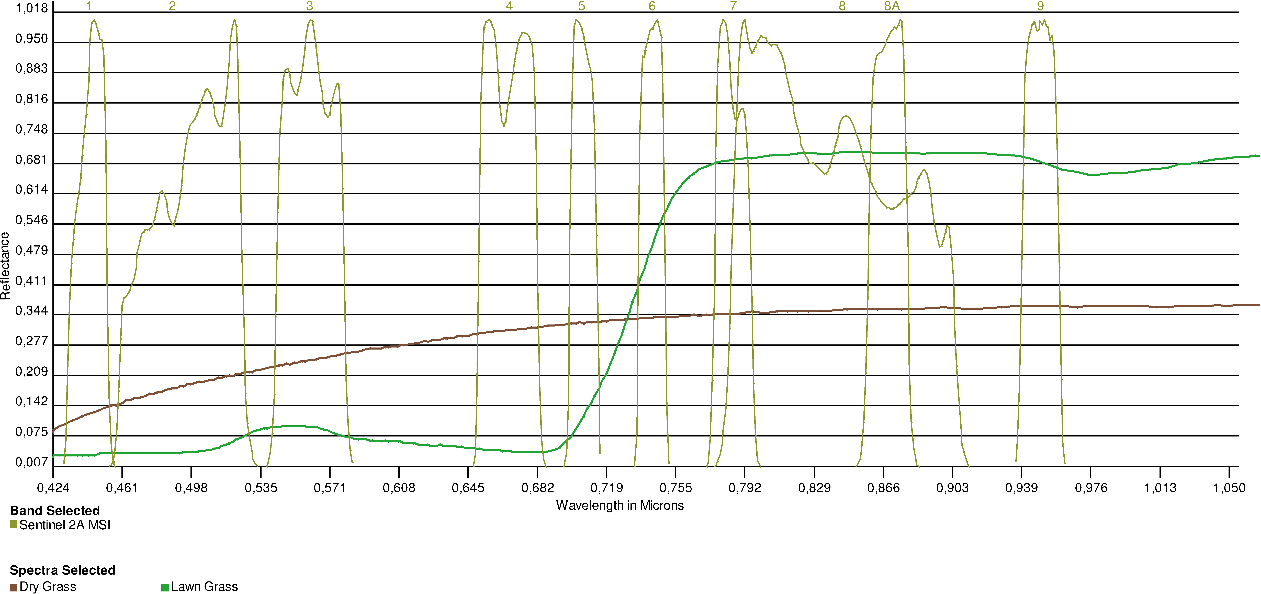}}
\caption{Reflective spectra of fresh (green curve) and dry (brown curve) grass in a wavelength range covered by the MSI/Sentinel-2 bands (yellow curves labelled at the top of the graph). There is a strong jump in the green grass spectrum between band 4 and band 7 (USGS Spectral Viewer, NASA, \capurl{http://landsat.usgs.gov/tools_viewer.php}).}
\label{f:refspec}
\end{figure*}

A proper choice of optical filters not only permits distinguishing between water and landscape, but also allows deciphering the state of vegetation or surface conditions. For instance, Fig.~\ref{f:refspec} indicates a noticeable difference between the reflective spectra of fresh (green curve) and dry (brown curve) grass. The main reason for this is the absorption power of 
\newglossaryentry{chlorophyll}{name=Chlorophyll,description={Chlorophyll is the pigment in green vegetation that is responsible for its colour. It is essential in photosynthesis, allowing plants to absorb energy from light.}}
chlorophyll. In particular, the transition between the red (band 4) and the infrared ranges (bands 7 to 9) sees a sudden jump in the spectrum of fresh, green grass, while the spectrum of dry grass remains rather constant. When subtracting the signals of the bands, one can distinguish between the two states.

\begin{figure}[!ht]
\centering
\resizebox{\hsize}{!}{\includegraphics{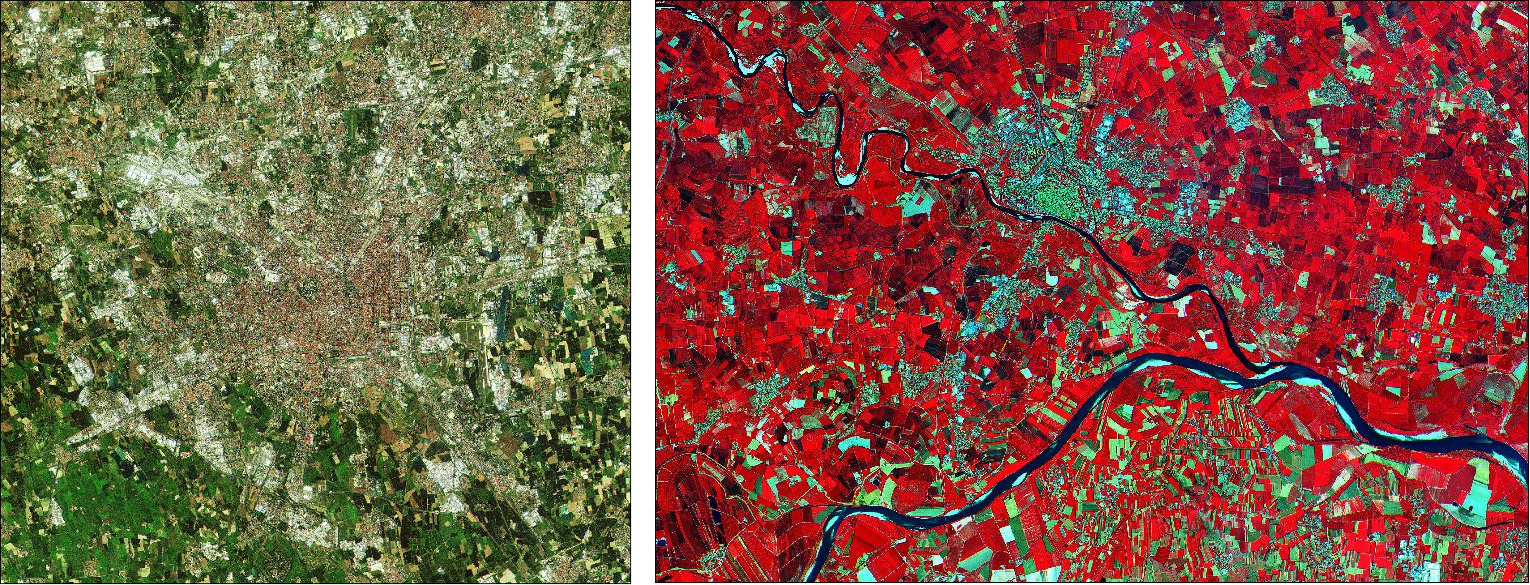}}
\caption{Images obtained with MSI/Sentinel-2A. Left: Realistic RGB coloured image of the city of Milan; right: false colour visualisation of the area around the river Po, Italy. The colour red represents the near infrared band which is sensitive for green vegetation (Copernicus data 2015/ESA, \capurl{https://directory.eoportal.org/web/eoportal/satellite-missions/c-missions/copernicus-sentinel-2}).}
\label{f:satimages}
\end{figure}

Satellite images contain
\newglossaryentry{pix}{name=Pixel,description={Smallest element digital images are made of. The name is derived from the two words {\em picture} and {\em element}. Each pixel possesses a value that represents the brightness of the portion of light that illuminates the part of the camara detector from which the image is produced.}}
pixel values that represent the brightness or intensity of the light reflected from the surface and detected in an optical band. They are usually displayed in grey-scale. Combining those images following to the rules of additive mixture of colour stimuli allows constructing colour images. When selecting the images of the spectral bands representing the colours red, green, and blue, the resulting RGB image displays the colours in a realistic way (Fig.~\ref{f:satimages}, left).

\subsection{Spectral index}
By merging data from different optical bands, much can be learnt about vegetation or construction areas in a qualitative way (Fig.~\ref{f:satimages}, right). If quantitative information is required, a more detailed analysis is needed. An established tool is a spectral index. This is a number that is calculated from data obtained at different wavelengths and allows comparing the relative brightness of different wavelengths of light that is reflected by the Earth’s surface.

\subsubsection{Normalised Differenced Vegetation Index (NDVI)}
An important spectral index used for identifying healthy vegetation is the {\em Normalised Differenced Vegetation Index (NDVI}). It is calculated from the measured intensities obtained in the red (R) and near infrared (NIR) spectral regimes. As mentioned, the transition between those bands is diagnostic in distinguishing between green vegetation from other features (Fig.~\ref{f:refspec}). It is calculated as follows.
\begin{equation}
NDVI = \frac{NIR-R}{NIR+R}
\end{equation}
With:

\medskip\noindent
\begin{tabular}{p{0.1\hsize}p{0.8\hsize}}
$R$:   & Intensity/brightness of reflected light in the red filter (ca.~0.6--0.7\,$\mu$m)\\
$NIR$: & Intensity/brightness of reflected light in the near infrared filter (ca.~0.8--0.9\,$\mu$m)\\
\end{tabular}

\begin{figure}
\centering
\resizebox{\hsize}{!}{\includegraphics{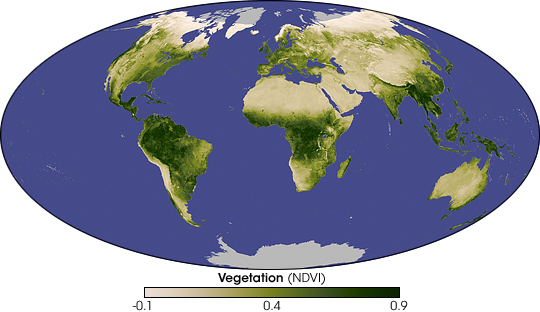}}
\caption{NDVI world map of November 2007 based on data of the ``Resolution Imaging Spectroradiometer (MODIS)'' of the NASA Terra satellite (NASA, \capurl{http://earthobservatory.nasa.gov/IOTD/view.php?id=8622}).}
\label{f:nvdiworld}
\end{figure}

\medskip
They are provided by the bands 4 and 8 of the Sentinel-2 MSI camera (Tab.~\ref{t:s2specbands}). The difference between $NIR$ and $R$ is normalised by their sum resulting in a range of values between $-1$ and $+1$. Negative values indicate water areas. A value between 0 and 0.2 represents nearly vegetation free surfaces, while a value close to $+1$ hints to a high coverage of green vegetation.

\subsubsection{Normalised Differenced Moisture Index (NDMI)}
Another spectral index is the {\em Normalised Differenced Moisture Index (NDMI)} or {|em Normalised Differenced Water Index (NDWI)}. It is sensitive for humid vegetation and open wetland. It supplements the NDVI.
\begin{equation}
NDMI = \frac{NIR-SWIR}{NIR+SWIR}
\end{equation}
With:

\medskip\noindent
\begin{tabular}{p{0.1\hsize}p{0.8\hsize}}
$NIR$:  & Intensity/brightness of reflected light in the near infrared filter (ca.~0.8--0.9\,$\mu$m)\\
$SWIR$: & Intensity/brightness of reflected light in the short-wave infrared filter (ca.~1.5--1.8\,$\mu$m)\\
\end{tabular}

\medskip
The $NDMI$ helps distinguishing between dry and wet areas.

\subsection{Modified Normalised Differenced Water Index (MNDWI)}
The {\em Modified Normalised Differenced Water Index (MNDWI)} is regarded as an improvement of the $NDMI$. It helps identifying open wetland and excludes artificial buildings, vegetation and agricultural areas.
\begin{equation}
MNDWI = \frac{G-SWIR}{G+SWIR}
\end{equation}
With:

\medskip\noindent
\begin{tabular}{p{0.1\hsize}p{0.8\hsize}}
$G$:   & Intensity/brightness of reflected light in the green filter (ca.~0.5--0.6\,$\mu$m)\\
$SWIR$: & Intensity/brightness of reflected light in the short-wave infrared filter (ca.~1.5--1.8\,$\mu$m)\\
\end{tabular}

\medskip
Open wetland attains higher positive values than with the $NDWI$, while other landmarks like buildings, vegetation and crop land have negative values.

\subsection{The software LEO Works 4}
The European Space Agency (ESA) has developed an educational tool for teaching and learning the basic steps of analysing satellite data. The latest version\footnote{\href{http://leoworks.terrasigna.com}{http://leoworks.terrasigna.com}} is being developed by Terrasigna\footnote{\href{http://www.terrasigna.com}{http://www.terrasigna.com}} in Romania. Since it is based on Java, it is independent of the running operating system. It will be used for mastering this activity.

\begin{figure}[b]
\centering
\resizebox{0.7\hsize}{!}{\includegraphics{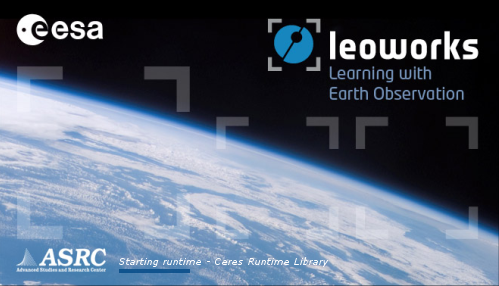}}
\caption{Launch window of LEO Works 4, a software for treating and analysing satellite data for educational purposes. It can be downloaded at \capurl{http://leoworks.terrasigna.com} and runs on a wide variety of operating systems.}
\label{f:leoworks}
\end{figure}

\section{List of material}
\begin{itemize}
\item Worksheet for students (needs to contain background information and activity steps)
\item Computer (the software needed is independent of the operating system)
\item Software installed: LEO Works 4, download at: \url{http://leoworks.terrasigna.com}
\item For the extension for advanced students:\\
Landsat satellite data files:\\
Venice\_Landsat\_ETM\_multispectral\_Jan2002.tif\\
Venice\_Landsat\_ETM\_multispectral\_Jul2002.tif
\end{itemize}

\section{Goals}
Students will get an insight into how multi-spectral satellite images can be diagnostic in deciphering Earth surface features like vegetation and the degree thereof as well as open water areas. They will get a hands-on understanding of how real remote sensing satellite data are being analysed. This will be done via a specially designed educational software package (LEO Works) which permits close to professional treatment of up to date satellite data. Students will understand the importance of such data for the lives of billions of people around the Earth and maybe grow interest in working in this field. Finally, the students will produce images and maps that are needed for the analysis. In the end, the students will be confident analysing satellite data on their own.

\section{Learning objectives}
\begin{itemize}
\item Students will inspect and analyse real satellite data at a close to professional standard.
\item Students will combine datasets to produce colour images and maps of spectral indices.
\item Students will answer questions and identify different surface features, such as vegetation and open water, by interpreting the maps of spectral indices.
\item Students will answer questions to discuss the importance of satellite data when dealing with issues like disaster management and climate change.
\end{itemize}

\section{Target group details}
Suggested age range: 14 -- 16 years\\
Suggested school level: Secondary School\\
Duration: 1.5 hours

\section{Evaluation}
The major part of this activity is analysing satellite images. The products created during this exercise are images generated by combining the images in a certain way. The success can be evaluated by comparing the maps and images with the ones provided with this material. In addition, students will answer questions that will show how well they understood the importance of satellite data for various aspects. These answers can be discussed as a class after the activity.

\section{Full description of the activity}
\subsection{Preparation}
Make printed or digital copies of the worksheet available to students. This contains the information in the background information which is needed to successfully analyse the data.

Install the LEO Works 4.0 software \url{http://leoworks.terrasigna.com} and make it available on the students' computers.  It is required to perform this activity.

\begin{figure}
\centering
\resizebox{\hsize}{!}{\includegraphics{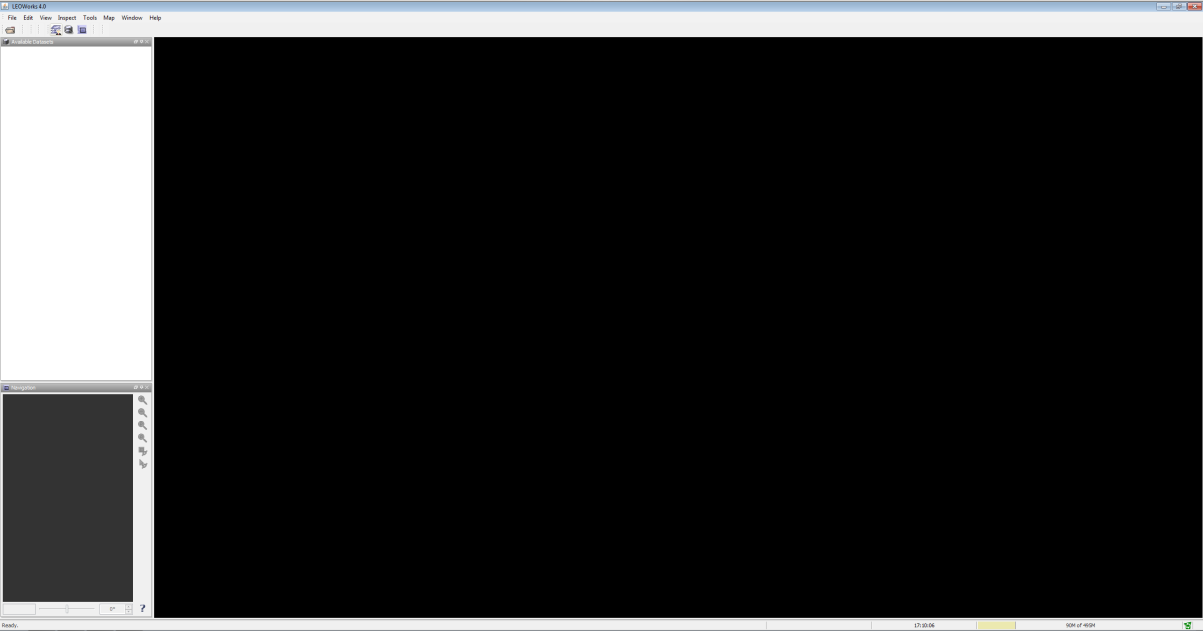}}
\caption{LEO Works 4 workspace. The menu bar contains procedures and tools for displaying and analysing the data. There are three windows below that provide a list of the loaded data sets and image displays.}
\label{f:lw:workspace}
\end{figure}

\subsection{Introduction}
Introduce the topic by asking students what they know about Earth observing. How can we observe the Earth and what is remote sensing? What information can we collect by remote sensing and what are their applications? The most obvious answers should include weather satellites.

Ask the students, if they knew where the images in Google Maps or Earth come from. The source of the images is mentioned at the bottom of the screen. They might find names like SPOT or Landsat. Ask students to choose one of these satellite campaigns to research on their background. Let them compile information on satellite launch dates, their orbits and countries of origin.

\subsection{Hands-on activity}
The activity is set up as a step-by-step instruction to analyse real satellite data. The exercise is interspersed with questions to evaluate the students' understanding as well as to point to the relevance of the satellite data. Some tasks contain very similar and repeating procedures that are used to reinforce the steps used in the analysis.

\subsection{Analysis of satellite imagery data using LEO Works 4}
This activity introduces basic tasks for processing and analysing remote sensing satellite data. 

The installed version already contains some example data sets that can be used for exercise purposes. They are stored in the {\em leoworks.data} folder. When using MS Windows, it can be found in the user directory. From the existing data sets, the one labelled Venice will be used.

\subsubsection{Reading the data}
\begin{figure}[!ht]
\centering
\resizebox{\hsize}{!}{\includegraphics{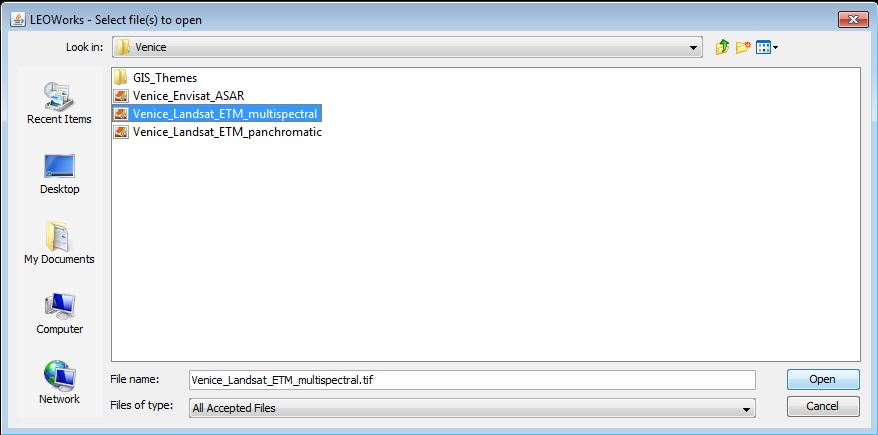}}
\caption{Window for file selection.}
\label{f:lw:files}
\end{figure}

After launching, the software presents its workspace as shown in Fig.~\ref{f:lw:workspace}. Open the file {\em Venice\_Landsat\_ETM\_multispectral.tif} by clicking on the first icon in the menu bar or via the menu {\em File $\rightarrow$ Open $\rightarrow$ Single File Dataset(s)}. A window appears from which the file is selected (Fig.~\ref{f:lw:files}).

The file contains seven individual images obtained in seven bands of the camera {\em Enhanced Thematic Mapper Plus (ETM+)} of NASA’s Landsat 7 satellite (Tab.~\ref{t:L7specbands}) covering the vicinity around the city of Venice in Italy. When the window {\em Specify Subset} appears, acknowledge by clicking OK.

\begin{table}[!ht]
\caption{List of the seven spectral bands of the {\em Enhanced Thematic Mapper Plus (ETM+)} camera of the Landsat~7 satellite (Source: NASA; column with colours is not revealed to students).}
\label{t:L7specbands}
\begin{tabular}{cccc}
\hline\hline
Landsat 7 & Wavelength & Resolution & Colour \\
          & ($\mu$m)   & (m)        &  \\
\hline
Band 1 &  0.450 --  0.515 & 30 & Blue \\
Band 2 &  0.525 --  0.605 & 30 & Green \\
Band 3 &  0.630 --  0.690 & 30 & Red \\
Band 4 &  0.750 --  0.900 & 30 & NIR \\
Band 5 &  1.550 --  1.750 & 30 & SWIR \\
Band 6 & 10.400 -- 12.500 & 60\tablefootmark{a} (30) & Thermal IR \\
Band 7 &  2.090 --  2.350 & 30 & IR \\
\hline
\end{tabular}
\tablefoottext{a}{\footnotesize The data were obtained with a spatial resolution of 60~m and scaled to a 30~m resolution.}
\end{table}

\begin{figure}[!ht]
\centering
\resizebox{\hsize}{!}{\includegraphics{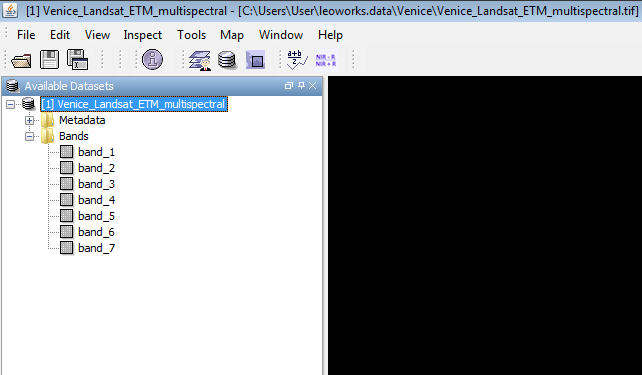}}
\caption{List of loaded data.}
\label{f:lw:datalist}
\end{figure}

The data automatically appear in the window to the upper left. The element {\em Bands} can be expanded by clicking on it to show the list of the seven images (Fig.~\ref{f:lw:datalist}). They are labelled band\_1 to band\_7 and correspond to the spectral bands of Tab.~\ref{t:L7specbands}.

\action{Fill in the column labelled {\em Colour} of Tab.~\ref{f:lw:datalist} for bands 1 to 5. Use the information provided with the introduction of the spectral indices}

\subsubsection{Image display}
A double-click on the band name issues a command that displays the image.

\action{Do this for band 1 first.}

You will see an image of the city of Venice and its surroundings. It consists of different shades of grey, a grey-scale display, that correspond to the brightness or intensity measured at a given spot (pixel) in the image. The contrast is a quite poor and should be adjusted using the tool {\em Interactive Stretching}.

\action{Find the corresponding button or menu item.}

\begin{figure}[!ht]
\centering
\resizebox{\hsize}{!}{\includegraphics{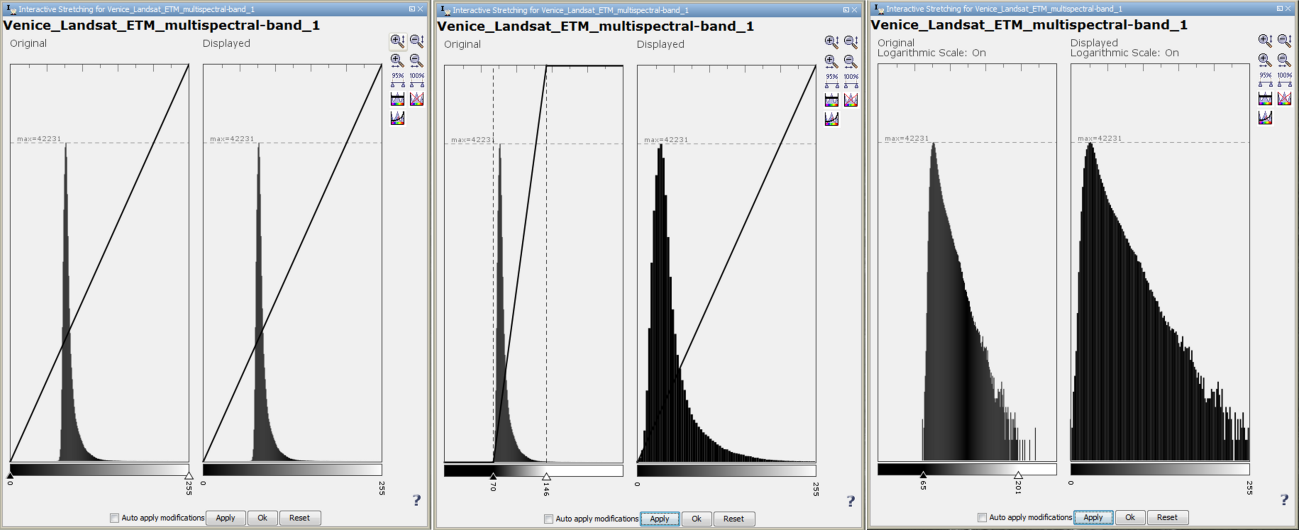}}
\caption{Windows for adjusting the contrast levels using {\em Interactive Stretching}. A window contains two graphs showing the distribution of pixel values in the image and the ones used for display, respectively. Adjustment is done by moving the flags. The setting is adopted by clicking {\em Apply}. Left: Distribution before adjustment; middle: after adopting the adjustment; right: the same shown in logarithmic scale, acquired by clicking the bottom left icon to the right.}
\label{f:lw:gscale}
\end{figure}

You can explore the meaning of the different buttons when moving the mouse pointer above them. After clicking, a new window appears as shown in Fig.~\ref{f:lw:gscale}.

The scaling of the contrast is accomplished by moving the flags. The window provides additional tools like displaying the data in a logarithmic scale.

\begin{figure}[!ht]
\centering
\resizebox{\hsize}{!}{\includegraphics{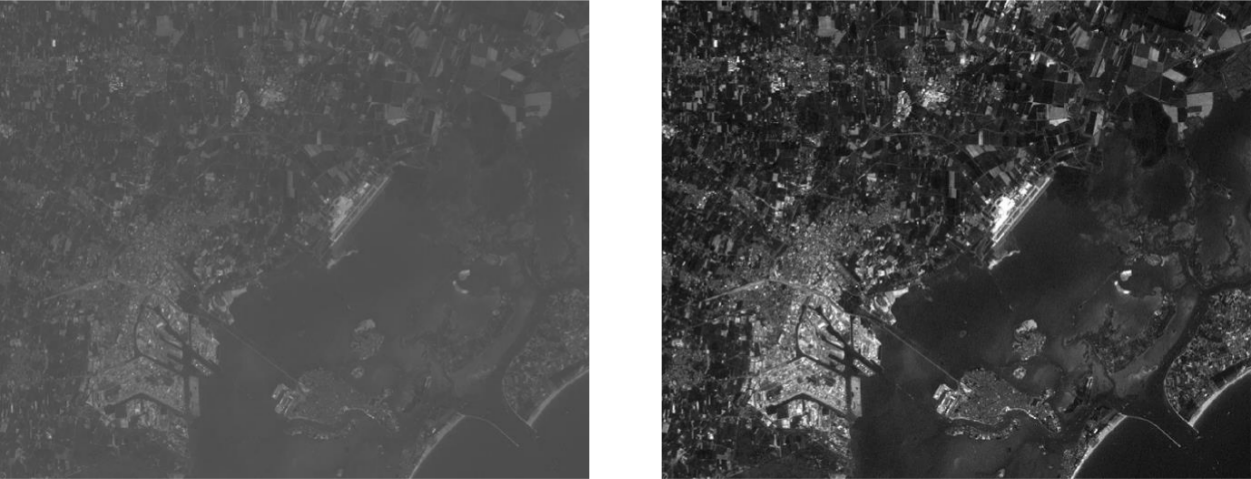}}
\caption{Image of band 1 before (left) and after (right) adjusting contrast scaling.}
\label{f:lw:contrast}
\end{figure}

\action{Display the seven images and adjust their scaling.}

\subsubsection{Creating a realistically coloured image}
After having adjusted the contrast settings, a colour picture can be produced by superposing three images. A bad contrast will lead to pale colours. For a realistic impression, the three bands representing blue, green, and red have to be selected.

\action{Find the corresponding bands in Tab.~\ref{t:L7specbands}. If you need help assigning colours to wavelengths, research the missing information on the internet.}

\begin{figure}[!ht]
\centering
\resizebox{\hsize}{!}{\includegraphics{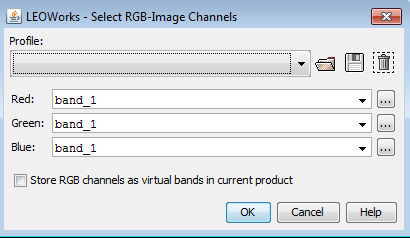}}
\caption{Window for selecting the bands to be used for constructing an RGB image.}
\label{f:lw:rgbchans}
\end{figure}

Select {\em View $\rightarrow$ New RGB View}. A new window appears (Fig.~\ref{f:lw:rgbchans}). Choose the matching bands for red, green, blue and click on OK.

\begin{figure}[!ht]
\centering
\resizebox{\hsize}{!}{\includegraphics{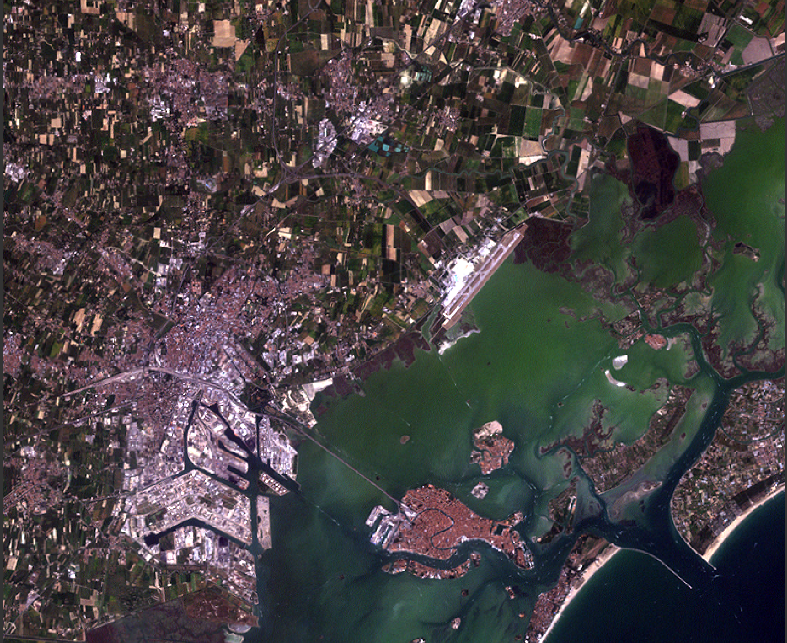}}
\caption{Three-colour image (RGB) created from satellite data of Venice.}
\label{f:lw:rgbimage}
\end{figure}

A new colour image appears (Fig.~\ref{f:lw:rgbimage}). If necessary, you can adjust the colours with {\em Interactive Stretching}.

\action{Inspect the result and try to identify landscape elements (buildings, water, soil, vegetation).}
\action{Find the airport.}

\subsubsection{Creating a false colour image}
You have just produced an RGB image that corresponds to the natural impression of colours how humans see it. It consists of the colours red, green, and blue. Imagine other species like bees or snakes. They can see other parts of the electromagnetic spectrum like the ultraviolet (UV) or the infrared (IR). We can simulate such kind of vision skills by combining different spectral bands than red, green and blue. The resulting colours do not match the natural ones we can see with our eyes, but they can help making interesting details visible.

Use the knowledge that the chlorophyll in green plants absorbs red light but reflects infrared radiation.

\action{Produce a three-colour image from the near infrared (ca.~0.8\,$\mu$m), red (ca.~0.65\,$\mu$m), and green (ca.~0.5\,$\mu$m). What are the corresponding bands? Put the infrared band in the red channel, the red band in the green channel and the green band in the blue channel of the RGB image.}
\action{Compare this image with Fig.~\ref{f:lw:rgbimage}. Where do you find green vegetation?}
\action{Can you distinguish between green crops and green water (algae)?}
\action{What does uncultivated land look like?}

\begin{figure}[!ht]
\centering
\resizebox{\hsize}{!}{\includegraphics{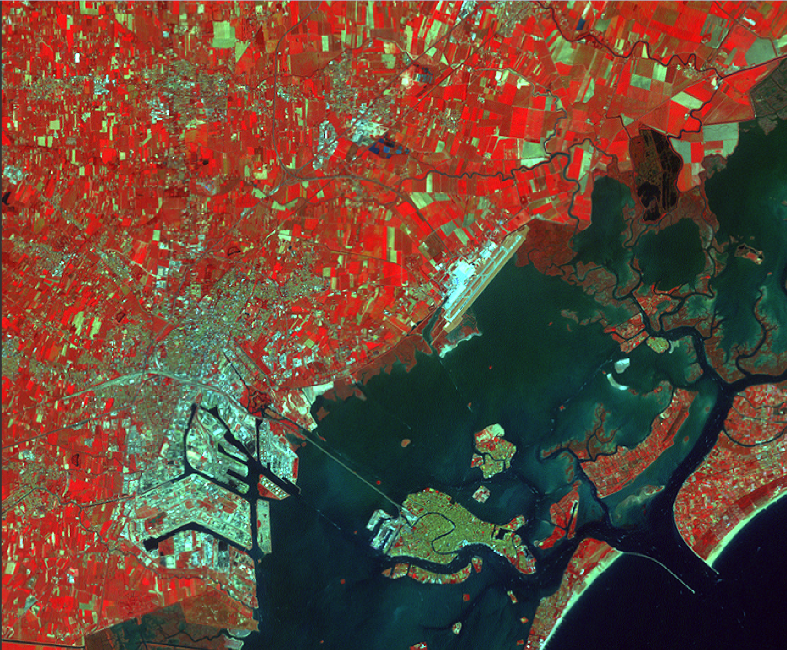}}
\caption{False colour image produced by combining the green, red and infrared bands.}
\label{f:lw:falseimage}
\end{figure}

\subsubsection{Analysis via NDVI}
You have already seen in the information section that the NDVI is a colour or spectral index
\begin{equation}
NDVI = \frac{NIR-R}{NIR+R}
\label{e:ndvi}
\end{equation}
that is particularly sensitive to green vegetation. The index provides a number that objectively reflects the degree of vegetation. Remember that there is a jump in the spectrum of green vegetation between the red (R) and the infrared (NIR) range (Fig.~\ref{f:refspec}). You will now construct a map that contains the NDVI for every image pixel. LEO Works provides a tool for this.

\action{Find the NDVI tool.}

\begin{figure}[!ht]
\centering
\resizebox{0.4\hsize}{!}{\includegraphics{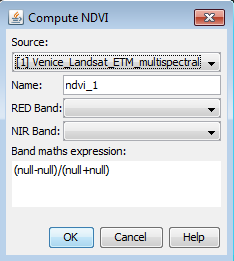}}
\caption{Window of the NDVI computing tool.}
\label{f:lw:ndvitool}
\end{figure}

After activating that tool, a new window pops up (Fig.~\ref{f:lw:ndvitool}. You select the dataset at the top. The next line contains the name of the image to be constructed and how it appears in the list of data. A name is already suggested. Select the suitable bands in the following rows below.

\action{What are the bands to be selected here? The answer can be found in the section about the NDVI and Tab.~\ref{t:L7specbands}.}

The formula is shown in Eq.\ref{e:ndvi}. In the beginning, the variables show {\em null} as long as no band is selected. It is automatically updated as soon as you select the band corresponding to the NIR and the R bands. The NDVI map is created by clicking {\em OK}. A suitable false colour representation is chosen automatically, which helps identify green vegetation. However, the scaling of the colour table must be adjusted.

\begin{figure}[!ht]
\centering
\resizebox{\hsize}{!}{\includegraphics{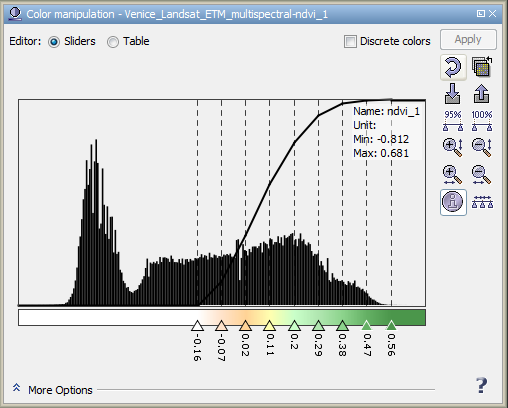}}
\caption{Window that allows adjusting the colour table.}
\label{f:lw:ndviscale}
\end{figure}

The tool {\em Color manipulation} is used for this. Move the flags of maximum value to the upper end of the distribution histogram. Then move the flag of the minimum value until the first green coloured flag reaches a value of 0.2 (Fig.~\ref{f:lw:ndviscale}). The new setting is adopted after clicking {\em Apply}.

The result should look similar to Figure 19. You see large white zones with alternating yellow and green areas in between.

\begin{figure}[!ht]
\centering
\resizebox{\hsize}{!}{\includegraphics{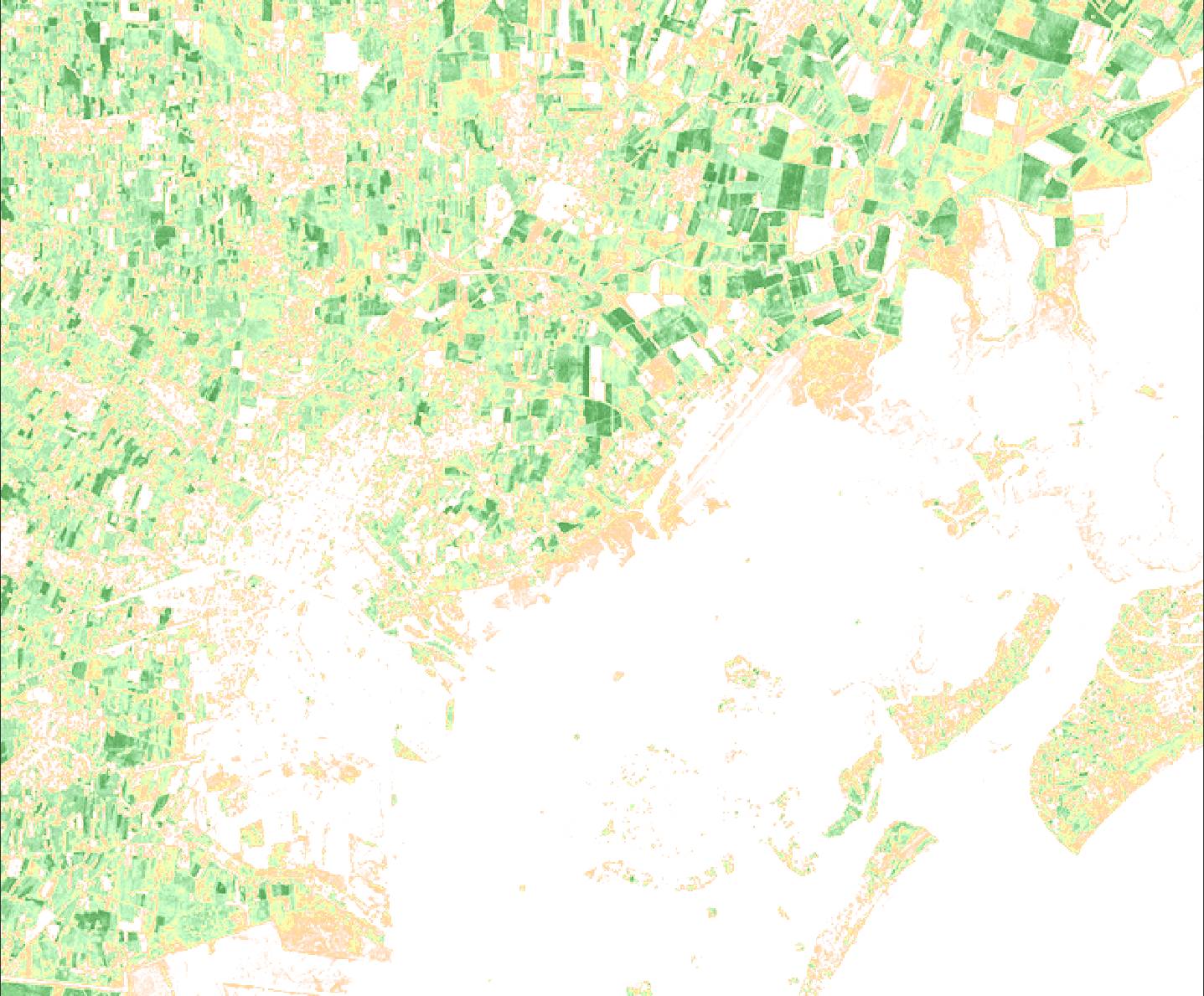}}
\caption{Map of the NDVI in the vicinity of Venice, based on Landsat 7 satellite data.}
\label{f:lw:ndvimap}
\end{figure}

\action{Compare the NDVI map with the previously produced images. What can you say about the degree of vegetation in the green and yellow areas?}
\action{Would you be able to detect a seasonal change, if the images were taken at a monthly rate?}
\action{What would be the situation during a draught?}

\subsection{Analysis via MNDWI}
You will now use the satellite data to identify open wetland with the MNDWI.
\begin{equation}
MNDWI = \frac{G-SWIR}{G+SWIR}
\end{equation}
Especially small ponds and narrow rivers are not easily found on naturally coloured images. The MNDWI can theoretically be constructed using the NDVI tool. However, the correct assignment of the corresponding bands can be confusing. LEO Works provides a generic tool to does all kinds of mathematical operations with the spectral bands. The procedure is called {\em Band arithmetic}.

\action{Find the tool in the tool bar or in the menu and open it.}

\begin{figure}[!ht]
\centering
\resizebox{\hsize}{!}{\includegraphics{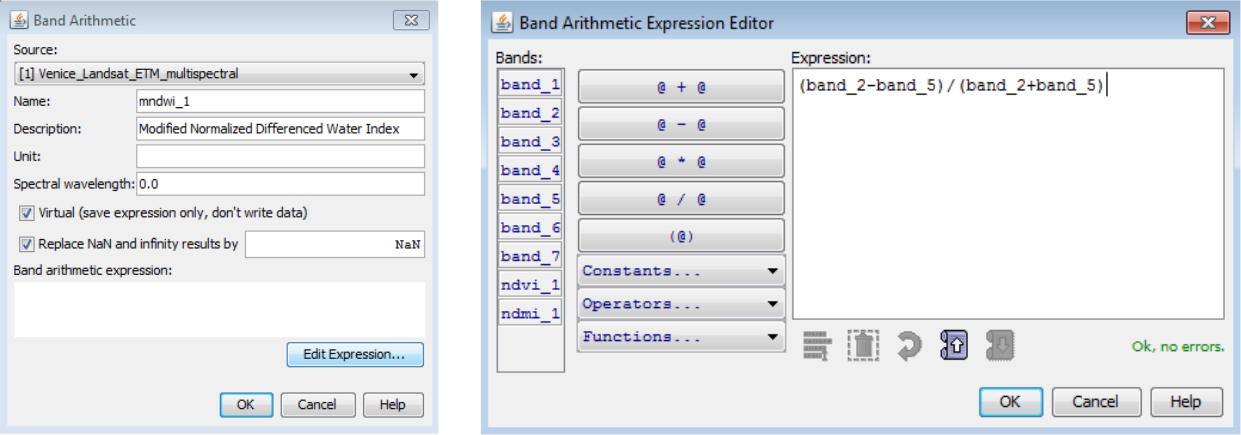}}
\caption{Windows for doing mathematical operations on the spectral band images.}
\label{f:lw:bandari}
\end{figure}

Similar to the tool for calculating the NDVI, you first select the dataset and the name of the image to construct (Fig.~\ref{f:lw:bandari}, left). Then click {\em Edit expression $\ldots$} for opening a new window (Fig.~\ref{f:lw:bandari}, right). This is where you enter the formula for calculating the spectral index.

\action{Find out what bands are needed to calculate the MNDWI.}

From the formula of this index you see that you divide the difference of the intensities of the reflected light measured in two spectral bands by their sum. Be careful with placing operators and brackets according to the formula.

After confirming the formula, it also appears in the first window. The procedure is executed by clicking {\em OK}.

\begin{figure}[!ht]
\centering
\resizebox{\hsize}{!}{\includegraphics{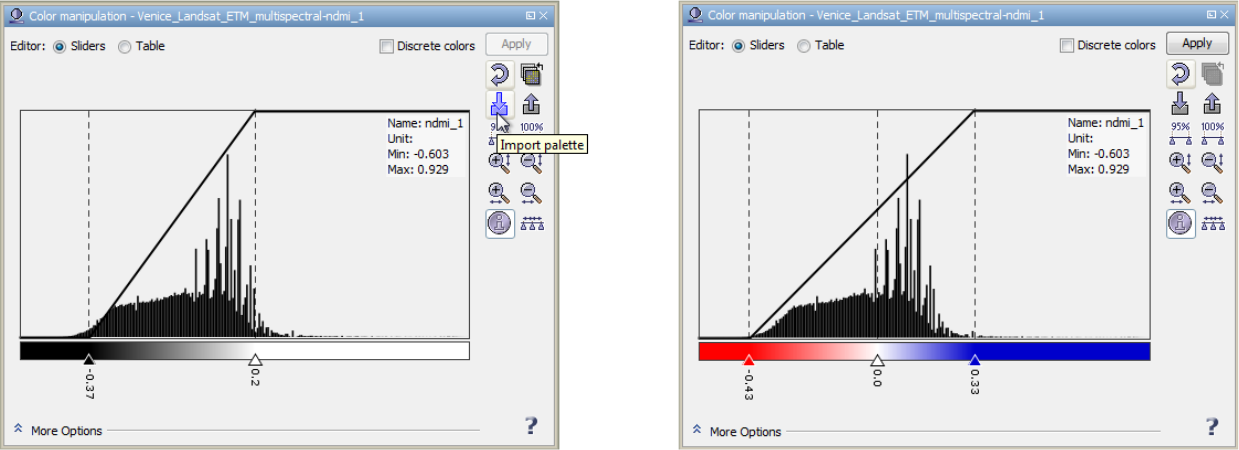}}
\caption{Colours can be assigned to image values to improve the readability of the map.}
\label{f:lw:palette}
\end{figure}

The resulting image presents the values of the index in grey-scale. To improve the readability of the map, you can change colours to certain values via the {\em Color manipulation} tool. A colour table is assigned by clicking on the symbol {\em Import palette} as shown in Fig.~\ref{f:lw:palette}.

\action{Select the file {\em gradient\_red\_white\_blue.cpd}.}
\action{Adjust the flags such that the values are well covered and the central flag represents the value 0.}
\action{What is the colour coding of water?}
\action{Compare the MNDWI map with the previously produced images. Would you be able to find wetland also on the naturally coloured image?}
\action{Can you imagine situations for which the identification of water levels can be important or even life-saving?}
\action{What would the image look like, if the water level rises?}
\action{If you have time, produce a map of the NDMI. Compare it with the other results.}

\begin{figure}[!ht]
\centering
\resizebox{\hsize}{!}{\includegraphics{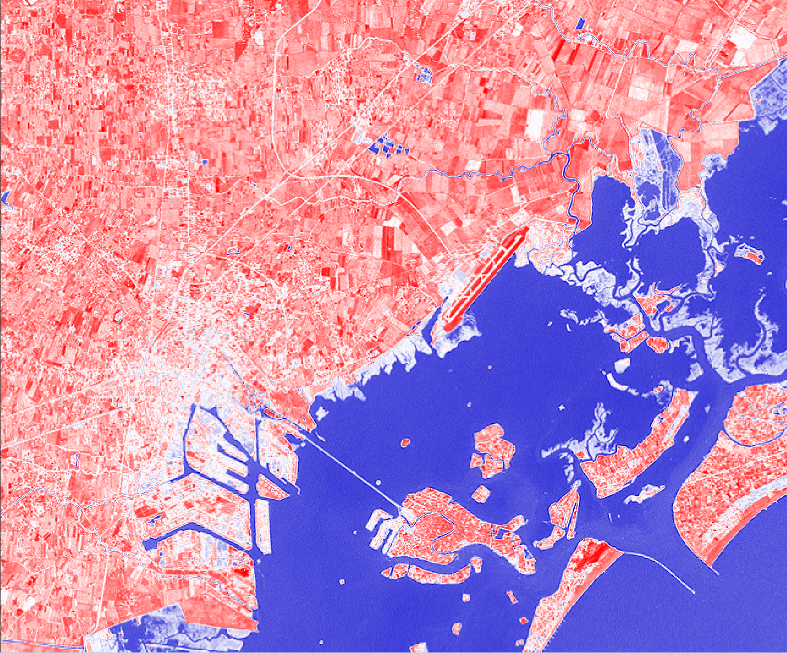}}
\caption{Map of the MNDWI in the vicinity of Venice based on Landsat 7 satellite data.}
\label{f:lw:mndwimap}
\end{figure}

\subsection{Additional activity for advanced students}
Two additional datasets are provided that show the same area in January and July 2002. The already analysed dataset is from August 2001.

\action{Load the two additional datasets like the previous one.}
\action{Produce naturally coloured RGB images.}
\action{Produce images of the NDVI distributions.}
\action{Compare the results from the three datasets obtained at different dates during the year. Indicate, how the vegetation changes.}
\action{In light of the results, describe and explain the advantage of satellite remote sensing.}

\section{Connection to school curriculum}
This activity is part of the Space Awareness category ``Our Fragile Planet'' and related to the curricula topics:
\begin{itemize}
\item Composition and Structure
\item Climate change
\item Surface
\item Satellites
\end{itemize}

The activity covers the aspects of the subject of Geo\-graphy in the UK curriculum as outlined in Tab.~\ref{t:UKcurr} at the end of this document. Table~\ref{t:BWcurr} contains details about how this activity is related to the school curriculum of the German federal state of Baden-W\"urttemberg \citep{bildungsplan_bw_2014}.

\begin{longtab}
\begin{longtable}{p{0.15\hsize}p{0.2\hsize}p{0.55\hsize}}
\caption{\label{t:UKcurr}Sections of the UK school curriculum of the subject geography related to the activity.}\\
\hline\hline
Level & Exam board & Section \\
\hline
\endfirsthead
\caption{continued.}\\
\hline\hline
Level & Exam board & Section \\
\hline
\endhead
\hline
\endfoot
KS3 & -- & Geographical skills and fieldwork\\
    &    & -- use Geographical Information Systems (GIS) to view, analyse and interpret places and data \\
\hline
GSCE (2016) & AQA & Skills 3.4.5: Use of qualitative and quantitative data from both primary and secondary sources to obtain, illustrate, communicate, interpret, analyse and evaluate geographical information. Including:\\
 & & -- geo-spatial data presented in a geographical information system (GIS) framework satellite imagery.\\
 & & \\
 & & Maps in association with photographs:\\
 & & -- be able to compare maps \\
 & & -- photographs: use and interpret ground, aerial and satellite photographs \\
 & & -- describe human and physical landscapes (landforms, natural vegetation, land-use and settlement) and geographical phenomena from photographs.\\
\hline
GSCE (2016) & Edexel & Cartographic skills \\
 & & -- describe and interpret geo-spatial data presented in a GIS framework framework (e.g. analysis of flood hazard using the interactive maps on the Environment Agency website) \\
\hline
GSCE (2016) & OCR A and B & Geographical skills\\
 & & 1.6. Describe, interpret and analyse geo-spatial data presented in a GIS framework.\\
 & & 4.1. Deconstruct, interpret, analyse and evaluate visual images including photographs, cartoons, pictures and diagrams. \\
\hline
GCSE & WJEC A and B (2016) & Cartographic skills \\
 & & 3.4 Describe and interpret geo-spatial data presented in a GIS framework. \\
\hline
AS/A level & AQA (2016) & 3.5.2.5 ICT skills \\
 & & -- Use of remotely sensed data (as described in Core skills).\\
 & & \\
 & & 3.5.1 Quantitative data: understanding of what makes data geographical and the geo-spatial technologies (e.g. GIS) that are used to collect, analyse and present geographical data \\
\end{longtable}
\end{longtab}

\begin{longtab}
\begin{longtable}{p{0.09\hsize}p{0.05\hsize}p{0.08\hsize}p{0.68\hsize}}
\caption{\label{t:BWcurr}Sections of the school curriculum of the German federal state of Baden-W\"urttemberg related to the activity.}\\
\hline\hline
School & Level & Subject & Section \\
\hline
\endfirsthead
\caption{continued.}\\
\hline\hline
School & Level & Subject & Section \\
\hline
\endhead
\hline
\endfoot
Sek~I, Gym & All & NWT & 2.1 Erkenntnisgewinnung und Forschen\\
 & & & 2. Bestimmungshilfen, Datenblätter, thematische Karten und Tabellen nutzen\\
 & & & 3. Informationen systematisieren, zusammenfassen und darstellen\\
 & & & 5. Messdaten mathematisch auswerten, beschreiben und interpretieren\\
 & & & 6. große Datenmengen (auch computergestützt) erfassen, verarbeiten und visualisieren\\
 & & & \\
 & & & 2.2 Entwicklung und Konstruktion\\
 & & & 7. die Funktionsweise technischer Systeme analysieren\\
 & & & \\
 & & & 2.3 Kommunikation und Organisation\\
 & & & 1. Fachbegriffe der Naturwissenschaften und der Technik verstehen und nutzen sowie Alltagsbegriffe in Fachsprache übertragen\\
 & & & 4. zeichnerische, symbolische und normorientierte Darstellungen analysieren, nutzen\\
 & & & \\
 & & & 2.4 Bedeutung und Bewertung\\
 & & & 2. das Zusammenwirken naturwissenschaftlicher Erkenntnisse und technischer Innovationen erläutern\\
 & & & 4. naturwissenschaftlich-technische Problemstellungen vor dem Hintergrund gesellschaftlicher und ökologischer Wechselwirkungen analysieren\\
\hline
Sek~I, Gym & 8--10 & NWT & 3.2.2 Energie und Mobilität\\
 & & & 3.2.2.1 Energie in Natur und Technik\\
 & & & (1) die Bedeutung der Sonne für das Leben auf der Erde\\
 & & & \\
 & & & 3.2.4 Informationsaufnahme und -verarbeitung\\
 & & & 3.2.4.1 Informationsaufnahme durch Sinne und Sensoren\\
 & & & (1) die Verwendungsmöglichkeiten von Sensoren beschreiben\\
 & & & \\
 & & & 3.2.4.2 Gewinnung und Auswertung von Daten\\
 & & & (2/3) Messdaten mithilfe von Software auswerten und darstellen\\
 & & & (4/5) raumbezogene Daten darstellen und nutzen\\
\hline
Gym & 8--10 & NWT & 3.2.4.3 Informationsverarbeitung\\
 & & & (1) Beispiele der analogen oder digitalen Informationscodierung aus Natur und Technik beschreiben\\
\hline
Sek I & All & Physik & 2.1 Erkenntnisgewinnung\\
 & & & 5. mathematische Zusammenhänge zwischen physikalischen Größen herstellen und überprüfen\\
\hline
Gym & All & Physik & 2.1 Erkenntnisgewinnung\\
 & & & 5. Messwerte auch digital erfassen und auswerten\\
 & & & 6. mathematische Zusammenhänge zwischen physikalischen Größen herstellen und überprüfen\\
\hline
Sek I, Gym, & All & Physik & 2.2 Kommunikation\\
OS Gem & & & 2. funktionale Zusammenhänge zwischen physikalischen Größen verbal beschreiben\\
 & & & 4. physikalische Vorgänge und technische Geräte beschreiben\\
 & & & 6. Sachinformationen und Messdaten aus einer Darstellungsform entnehmen und in andere Darstellungsformen überführen\\
\hline
Sek I, Gym & 7--10 & Physik & 3.2.1 Denk- und Arbeitsweisen der Physik\\
 & & & (1) Kriterien für die Unterscheidung zwischen Beobachtung und Erklärung\\
\hline
Sek I, Gym & 7--9 & Physik & 3.2.2 Optik und Akustik\\
 & & & (7) Streuung und Absorption\\
 & & & (8) Reflexion an ebenen Flächen\\
 & & & (12) einfache Experimente zur Zerlegung von weißem Licht\\
\hline
Sek I, Gym & 10 & Physik & 3.3.3 Wärmelehre\\
 & & & (8) Auswirkungen des Treibhauseffektes auf die Klimaentwicklung\\
\hline
OS Gem & 11 & Physik & 3.2.1 Denk- und Arbeitsweisen der Physik\\
 & & & (1) Kriterien für die Unterscheidung zwischen Beobachtung und Erklärung\\
 & & & \\
 & & & 3.3.3 Wärmelehre\\
 & & & (8) Auswirkungen des Treibhauseffektes auf die Klimaentwicklung\\
\hline
Sek I, Gym, & All & Geografie & 2.1 Orientierungskompetenz \\
OS Gem & & & 1. geographische Sachverhalte in topografische Raster einordnen\\
 & & & 2. geographische Sachverhalte raum-zeitlich einordnen\\
 & & & \\
 & & & 2.2 Analysekompetenz\\
 & & & 1. geographische Strukturen und Prozesse herausarbeiten, analysieren und charakterisieren\\
 & & & 2. systemische Zusammenhänge darstellen und daraus resultierende zukünftige Entwicklungen erörtern\\
 & & & \\
 & & & 2.3 Urteilskompetenz\\
 & & & 1. geographisch relevante Beurteilungskriterien erläutern\\
 & & & 4. raumrelevante systemische Strukturen und Prozesse auch hinsichtlich ihrer zukünftigen Entwicklung bewerten\\
 & & & \\
 & & & 2.5 Methodenkompetenz\\
 & & & 1. fragengeleitete Raumanalysen durchführen\\
 & & & 2. Informationsmaterialien  in analoger und digitaler Form unter geographischen Fragestellungen problem-, sach- und zielgemäß kritisch analysieren \\
\hline
Sek I, Gym & 5--6 & Geografie & 3.1.1 Teilsystem Erdoberfläche\\
 & & & 3.1.1.1 Grundlagen der Orientierung\\
 & & & (4) die Nutzung analoger und digitaler Hilfsmittel zur Orientierung erläutern\\
 & & & \\
 & & & 3.1.5 Natur- und Kulturräume\\
 & & & 3.1.5.1 Analyse ausgewählter Räume in Deutschland und Europa\\
 & & & (1) die naturräumliche Gliederung Baden-Württembergs, Deutschlands und Europas beschreiben\\
\hline
Sek I,Gym & 7--9 & Geografie & 3.2.2 Teilsystem Wetter und Klima\\
 & & & 3.2.2.3 Phänomene des Klimawandels\\
 & & & (3) globale Auswirkungen des Klimawandels im Überblick beschreiben\\
\hline
Sek I, Gym, & 9--11 & Geografie & 3.3.1 Teilsystem Erdoberfläche\\
OS Gem & & & 3.3.1.1 Digitale Orientierung\\
 & & & (1) mithilfe von Informationen aus der Fernerkundung und aus Web-GIS Räume analysieren\\
\hline
Gym, & 11-13 & Geografie & 3.4.2/3.5.3 Globale Herausforderungen\\
OS Gem & & & 3.4.2.1/3.5.3.1 Globale Herausforderungen und Zukunftssicherung\\
 & & & (1) „Globale Herausforderungen“ charakterisieren\\
 & & & \\
 & & & 3.4.2.2/3.5.3.2 Globale Herausforderung: Klimawandel\\
 & & & (1) Ursachen und Dimensionen des Klimawandels auf der Grundlage aktueller wissenschaftlicher Erkenntnisse erläutern\\
\end{longtable}
\end{longtab}

\section{Conclusion}
The students used the LEO Works software to inspect and analyse real satellite data at a close to professional standard. They combined datasets to produce colour images and maps of spectral indices and learnt how to interpret them. Students should understand the importance of satellite data when dealing with issues like disaster management and climate change.

\begin{acknowledgements}
This resource was developed in the framework of Space Awareness. Space Awareness is funded by the European Commission's Horizon 2020 Programme under grant agreement no. 638653.
\end{acknowledgements}

%-------------------------------------------------------------------
\bibliographystyle{aa}
\bibliography{FragilePlanet}

\glsaddall
\printglossaries

\begin{appendix}
\section{Relation to other educational materials}
This unit is part of a larger educational package called ``Our Fragile Planet'' that introduces several historical and modern techniques used for navigation. An overview is provided via: \href{https://drive.google.com/file/d/0Bzo1-KZyHftXSnY4NGRGOGtodkE/view?usp=sharing}{Our\_Fragile\_Planet.pdf}

\section{Supplemental material}
The supplemental material is available on-line via the Space Awareness project website at \url{http://www.space-awareness.org}. The direct download links are listed as follows:
 
\begin{itemize}
 \item Worksheets: \href{https://drive.google.com/file/d/0Bzo1-KZyHftXZG1rVUQ4TGRjaEk/view?usp=sharing}{astroedu1618-A\_View\_From\_Above-WS.pdf}
 \item Additional dataset from January 2002: \href{https://drive.google.com/file/d/0Bzo1-KZyHftXOUR4ZWpYWE1rM2M/view?usp=sharing}{Venice\_Landsat\_ETM\_multispectral\_Jan2002.tif}
\item Additional dataset from July 2002: \href{https://drive.google.com/file/d/0Bzo1-KZyHftXQXFOc0syZmdvRkE/view?usp=sharing}{Venice\_Landsat\_ETM\_multispectral\_Jul2002.tif}
\end{itemize}

Suitable image material from other areas can be downloaded via the ESA Eduspace image server at:
\url{http://www.esa.int/SPECIALS/Eduspace_EN/SEMLK0F1EHH_0.html}

Another source of suitable satellite data:
\url{https://earthexplorer.usgs.gov}
\end{appendix}
\end{document}